\documentclass[12pt]{article}
\usepackage{amsmath,amsbsy,amssymb,amsthm,amsxtra,overpic,bbm,bm,epsfig,subfigure}
\usepackage{hyperref}
\usepackage{mathrsfs}
\usepackage{color}
\usepackage{comment}
\usepackage{float}
\usepackage{enumitem}
\usepackage{cite}

\usepackage{slashed,stmaryrd}

\def\thefootnote{\fnsymbol{footnote}}

\begin{document}
	
\vspace{0.2cm}
\begin{center}
{\Large \bf Testing unitarity of the $3\times 3$ neutrino mixing matrix \\
in an atomic system}
\end{center}

\vspace{0.2cm}

\begin{center}
	\small
	{\bf Guo-yuan Huang$^{a,b}$}\footnote{Email: huanggy@ihep.ac.cn},
	{\bf Noboru Sasao$^{c}$}\footnote{Email: sasao@okayama-u.ac.jp},
	~ {\bf Zhi-zhong Xing$^{a,b,d}$\footnote{Email: xingzz@ihep.ac.cn}},
	{\bf Motohiko Yoshimura$^{c}$\footnote{Email: yoshim@okayama-u.ac.jp}} \\
	{\small $^a$Institute of High Energy Physics, Chinese Academy of
		Sciences, Beijing 100049, China \\	$^b$School of Physical Sciences,
		University of Chinese Academy of Sciences, Beijing 100049, China \\
		$^c$Research Institute for Interdisciplinary Science, Okayama University,
Okayama 700-8530, Japan \\
		$^d$Center for High Energy Physics, Peking University, Beijing
		100080, China}
\end{center}

\vspace{1.5cm}

\begin{abstract}
Unitarity of the $3\times 3$ lepton flavor mixing matrix $V$ is unavoidably violated
in a seesaw mechanism if its new heavy degrees of freedom are slightly mixed with
the active neutrino flavors. We propose to use the atomic transition process
${\left| \rm e \right> \to \left|\rm g \right> +  \gamma + \nu^{}_{i} +
\overline{\nu}^{}_{j}}$ (for $i, j = 1, 2, 3$), where $\left|\rm e \right>$ and
$\left|\rm  g \right>$ stand respectively for the excited and ground levels of an
atomic system, to probe or constrain the unitarity-violating effects of $V$. We
find that the photon spectrum of this transition will be distorted by the
effects of $V V^\dagger \neq {\bf 1}$ and $V^\dagger V \neq {\bf 1}$ as compared
with the $V V^\dagger = V^\dagger V = {\bf 1}$ case. We locate certain frequencies
in the photon spectrum to minimize the degeneracy of effects of the unitarity
violation and uncertainties of the flavor mixing parameters themselves. The
requirements of a nominal experimental setup to test the unitarity of $V$ are
briefly discussed.
\end{abstract}

\def\thefootnote{\arabic{footnote}}
\setcounter{footnote}{0}

\newpage
\section{Introduction}

To naturally interpret the finite but tiny masses of three known neutrinos $\nu^{}_i$
(for $i =1,2,3$) corresponding to their flavor eigenstates $\nu^{}_\alpha$
(for $\alpha = e, \mu, \tau$), the most popular and well-motivated way is to
extend the standard electroweak model by
introducing three heavy sterile neutrinos and allow for lepton number violation
--- the canonical seesaw mechanism \cite{Minkowski:1977sc,Yanagida:1979as,
GellMann:1980vs,Glashow:1979nm,Mohapatra:1979ia}. In this connection the small
mixing between light and heavy degrees of freedom unavoidably gives rise to
a slight departure of the $3\times 3$ Pontecorvo-Maki-Nakagawa-Sakata (PMNS)
lepton flavor mixing matrix $V$ from unitarity (i.e., $V^{\dagger} V \neq \bm{1}$
and $ V V^{\dagger} \neq \bm{1}$) \cite{Antusch:2006vwa}, and it is particularly
appreciable in some interesting and testable TeV-scale seesaw models \cite{Xing:2009in}.
This kind of {\it indirect} unitarity violation of $V$ can in principle be probed or
constrained at low energies, such as in long-baseline accelerator neutrino
oscillation experiments \cite{FernandezMartinez:2007ms,Xing:2007zj,Goswami:2008mi,Luo:2008vp,
Li:2015oal,Tang:2017khg,Dutta:2019hmb} and medium-baseline reactor antineutrino
oscillation experiments \cite{Xing:2012kh,Qian:2013ora,Fong:2016yyh,Li:2018jgd}.

Different from previous works, the present paper aims to put forward a new and
interdisciplinary method for testing unitarity of the PMNS matrix $V$ by looking
at possible indirect unitarity violation of $V$. Our
approach is closely related to atomic physics and has nothing to do with neutrino or
antineutrino oscillations. Let us elaborate this novel idea in the following.

\begin{figure}[t!]
	\begin{center}
		\subfigure{
			\hspace{-0cm}
			\includegraphics[width=0.42\textwidth]{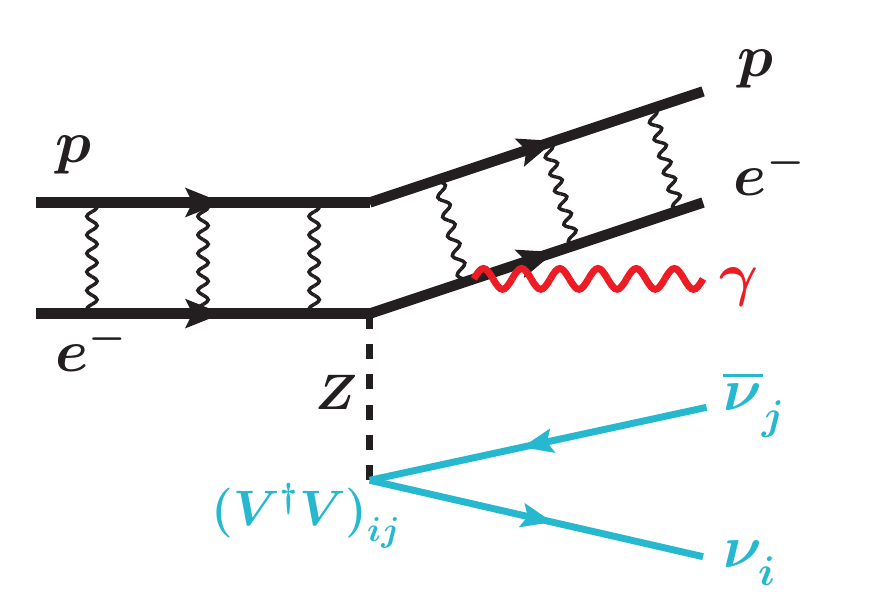} }
		\subfigure{
			\hspace{-0cm}
			\includegraphics[width=0.42\textwidth]{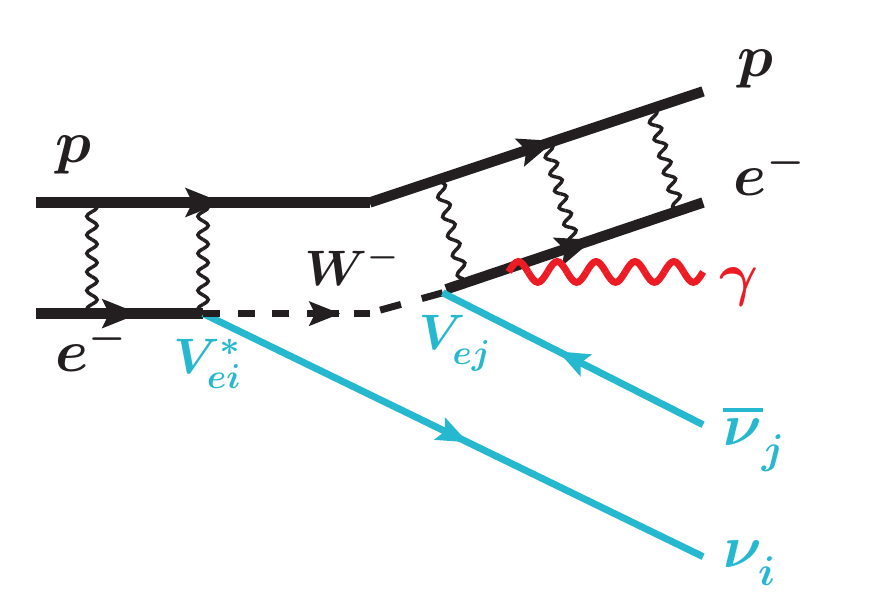} }
	\end{center}
	\vspace{-0.8cm}
	\caption{The Feynman diagrams for weak neutral- and charged-current
		interactions that contribute to the radiative emission of neutrino pairs in an
		atomic system.}
	\label{fig:sd}
\end{figure}

In 2006 one of us (M.Y.) proposed to use some fine atomic transitions as a powerful tool
to determine the absolute neutrino masses and the nature of massive neutrinos (namely,
whether they are the Majorana or Dirac particles) \cite{Yoshimura:2006nd}.
The relevant transition process in a feasible experimental setup is
${\left| \rm e \right> \to \left|\rm g \right> +  \gamma + \nu^{}_{i} +
\overline{\nu}^{}_{j} }$ (for $i, j = 1, 2, 3$),
where $\left|\rm e \right>$ is the excited level in an atomic or molecular system,
and $\left|\rm  g \right>$ denotes the ground one. Such a transition can take place
via an intermediate state $\left|\rm  v \right>$.
The information about neutrino properties is encoded in the spectrum
of the emitted photons $\gamma$, just like the spectrum of the emitted
electrons in a nuclear $\beta$-decay experiment. Before and after
the transition, the total energy of the system is conserved:
$E^{}_{\rm eg} = \omega + E^{}_{i}+E^{}_{j}$, where $E^{}_{\rm eg}$ represents the
energy difference between $\left| \rm e \right>$ and $\left| \rm g \right>$,
$\omega$ stands for the energy of the emitted photon, and $E^{}_{i}$ (or $E^{}_{j}$)
denotes the energy of the neutrino $\nu^{}_{i}$ (or $\overline{\nu}^{}_{j}$) with
the mass $m^{}_{i}$ (or $m^{}_{j}$). The Feynman diagrams responsible for the
transition under consideration are shown in Fig.~\ref{fig:sd}, where the relevant weak
neutral- and charged-current interactions are described by
\begin{eqnarray} \label{eq:firstLagrangian}
-\mathcal{L}^{}_{\rm nc} & = & \frac{g^{}_{\rm w}}{4 \cos{\theta^{}_{\rm w}}} \left[
\overline{\left(\begin{matrix} \nu^{}_{1} & \nu^{}_{2} & \nu^{}_{3}
	\end{matrix} \right) } \ \gamma^{\mu} \left(1- \gamma^{}_{5}\right) V^{\dagger} V \left(\begin{matrix} \nu^{}_{1} \\ \nu^{}_{2} \\ \nu^{}_{3} \end{matrix} \right)
Z^{}_{\mu}  + \overline{e} \ \gamma^{\mu} \left(4\sin^2{\theta^{}_{\rm w}} - 1 +
\gamma^{}_{5}\right) e Z^{}_{\mu} \right] \; ,
\nonumber \\
-\mathcal{L}^{}_{\rm cc} & = & \frac{g^{}_{\rm w}}{2\sqrt{2}}
\overline{\left(\begin{matrix} e & \mu & \tau \end{matrix} \right)}
\ \gamma^{\mu} \left(1- \gamma^{}_{5}\right) V \left(\begin{matrix} \nu^{}_{1}
\\ \nu^{}_{2} \\ \nu^{}_{3} \end{matrix}\right) W^{-}_{\mu} + {\rm h.c.} \; ,
\end{eqnarray}
where $g^{}_{\rm w}$ denotes the weak-interaction coupling constant, and
$\theta^{}_{\rm w}$ is the weak mixing angle. Integrating out the relevant heavy
degrees of freedom (i.e., the massive $W^\pm$ and $Z$ bosons)
in Eq.~(\ref{eq:firstLagrangian}) and performing the Fierz transformations,
we are left with the following effective four-fermion interactions at low energies:
\begin{eqnarray} \label{eq:Leff}
-\mathcal{L}^{}_{\rm eff} & = &
\frac{G^{}_{\rm F}}{2\sqrt{2}} \sum^3_{i=1}\sum^3_{j=1} \left[\overline{\nu}^{}_{i}
\gamma^{\mu} \left(1 -\gamma^{}_{5}\right) (V^{\dagger}V)^{}_{ij} \nu^{}_{j} \right]
\cdot \left[\overline{e} \ \gamma^{}_{\mu} \left(4\sin^2{\theta^{}_{\rm W}} - 1 +
\gamma^{}_{5}\right) e \right] \hspace{0.2cm}
\nonumber \\
&& + \frac{G^{}_{\rm F}}{\sqrt{2}} \sum^3_{i=1} \sum^3_{j=1} \left[
\overline{\nu}^{}_{i} \gamma^{\mu} \left(1 -\gamma^{}_{5}\right)
V^*_{ei} V^{}_{ej} \nu^{}_{j} \right] \cdot \left[\overline{e} \ \gamma^{}_{\mu}
\left(1-\gamma^{}_{5}\right) e\right] \; ,
\end{eqnarray}
where $G^{}_{\rm F} = g^2_{\rm w} / (4\sqrt{2} M^2_{W})\simeq 1.166
\times 10^{-5}~{\rm GeV^{-2}}$ is the Fermi constant. We assume that the
level transition associated with the neutrino-pair emission
is of the ${\rm M1}$ type via the electron spin flipping. The axial
current of the electron field will therefore dominate the transition,
as the contribution of the vector current in this case is suppressed by the
velocity of the nonrelativistic electron. As a result, Eq.~(\ref{eq:Leff})
is simplified to
\begin{eqnarray} \label{eq:Leff2}
\mathcal{L}^{}_{\rm eff} \longrightarrow
\mathcal{L}^{(\rm A)}_{\rm eff} = \frac{G^{}_{\rm F}}{\sqrt{2}} \sum^3_{i=1}\sum^3_{j=1}
\left\{\overline{\nu}^{}_{i} \gamma^{\mu}_{} (1 -\gamma^{}_{5}) \left[ V^*_{ei} V^{}_{ej}
- \frac{1}{2} (V^{\dagger}V)^{}_{ij}\right]
\nu^{}_{j} \cdot \overline{e} \ \gamma^{}_{\mu} \gamma^{}_{5} \ e \right\}  \; .
\end{eqnarray}
The axial electron current projected into the atomic levels
$\left< \rm v\right| \overline{e} \gamma^{\mu} \gamma^{}_{5} e \left|\rm e \right>$
can be reduced to $\left< \rm v\right| 2 \bm{S} \left|\rm e \right>$ in the
nonrelativistic limit, where $\bm{S}$ denotes the spin operator. Note that the
PMNS matrix $V$ is involved into the transition process as a single factor
$a^{}_{ij} \equiv V^*_{ei} V^{}_{ej} - (V^{\dagger}V)^{}_{ij}/2$,
which can be simplified to $V^{*}_{ei} V^{}_{ej} - \delta^{}_{ij}/2$ if $V$ is
exactly unitary.

A rough but instructive estimate based on the naive dimensional analysis yields
the transition rate $\Gamma \sim N^{}_{\rm tar} G^{2}_{\rm F}
E^5 \sim 10^{-8}~{\rm s^{-1}}$
for a macroscopic atomic ensemble with the target atom number $N^{}_{\rm tar}\sim\mathcal{O}(10^{23})$,
where $E \sim \mathcal{O}(1~{\rm eV})$ stands for
a typical energy transfer of the atomic level. That is why the transition rate
demands some magnification mechanisms for a realistic measurement.
In Refs.~\cite{Yoshimura:2008ya} and \cite{Yoshimura:2012tm}
it was proposed to utilize the
super-radiance (SR) phenomenon \cite{Dicke:1954zz} in quantum optics to enhance
the rate. The total transition rate for a macroscopic ensemble in the stochastic
case is just proportional to the number of total target atoms $N^{}_{\rm tar}$.
If atoms in the ensemble are arranged to behave collectively, however, the final
rate will be instead proportional to $N^2_{\rm tar}$. This coherence enhancement
makes a realistic observation possible. In this connection a review
of the radiative emission of neutrino pairs (RENP) has been done
by the SPAN (SPectroscopy with Atomic Neutrino) group in Ref.~\cite{Fukumi:2012rn}.

To achieve a coherence among the macroscopic target atoms,
the momenta of the outgoing particles must follow the relation
$\bm{p}^{}_{\rm eg} = \bm{k} + \bm{p}^{}_{i} + \bm{p}^{}_{j}$, where
$\bm{p}^{}_{\rm eg}$ denotes the initial phase imprinted on the medium
which can be manufactured to be nonzero by the coherence-establishing procedure
in the scenario of the boosted RENP \cite{Tanaka:2017juo},
$\bm{k}$ is the momentum of the photon,
and $\bm{p}^{}_{i}$ (or $\bm{p}^{}_{j}$) represents the momentum of the neutrino
$\nu^{}_{i}$ (or $\overline{\nu}^{}_{j}$). Therefore, both energy and momentum
conservations should be imposed on the system to have a successful coherent
enhancement. Instead of going into the detail in this aspect, we subsequently focus
on the particle-physics part of the RENP and illustrate how the effects of indirect
unitarity violation of $V$ can manifest them in this interesting process.

\section{Methodology}

A slight departure of the PMNS neutrino mixing matrix $V$ from unitarity can in
general be parametrized in the following way:
\begin{eqnarray} \label{eq:epsilon}
VV^{\dagger}_{} - \bm{1} =
\left(\begin{matrix} \epsilon^{}_{ee} & \epsilon^{}_{e\mu} & \epsilon^{}_{e\tau} \\
\epsilon^{}_{\mu e} & \epsilon^{}_{\mu \mu} & \epsilon^{}_{\mu \tau} \\
\epsilon^{}_{\tau e} & \epsilon^{}_{\tau \mu} & \epsilon^{}_{\tau \tau}
\end{matrix}\right) \; , \quad
V^{\dagger}V -\bm{1}  =
\left(\begin{matrix} \tilde{\epsilon}^{}_{11} & \tilde{\epsilon}^{}_{12} &
\tilde{\epsilon}^{}_{13} \\
\tilde{\epsilon}^{}_{21} & \tilde{\epsilon}^{}_{22} & \tilde{\epsilon}^{}_{23} \\ \tilde{\epsilon}^{}_{31} &
\tilde{\epsilon}^{}_{32} & \tilde{\epsilon}^{}_{33} \end{matrix}\right) \; ,
\end{eqnarray}
where $\epsilon^{}_{\alpha\beta}=\epsilon^{*}_{\beta\alpha}$
(for $\alpha, \beta = e, \mu, \tau$)
and $\tilde{\epsilon}^{}_{ij} = \tilde{\epsilon}^{*}_{ji}$ (for $i,j = 1,2,3$)
are two sets of small
unitarity-violating parameters, and their indirect correlation can be established
when a full parametrization of the $6\times 6$ flavor mixing matrix between three
species of light Majorana neutrinos and three species of heavy Majorana neutrinos
is made \cite{Xing:2011ur}. To see this point in an empirical way, one
may simply parametrize $V$ as $V = \left(1-\eta\right) U$, where $U$ is
exactly unitary and $\eta$ is Hermitian and its matrix elements are all small in
magnitude. Then $\epsilon^{}_{\alpha\beta} \simeq -2\eta^{}_{\alpha\beta}$ and
$\tilde{\epsilon}^{}_{ij} \simeq -2 (U^\dagger \eta U)^{}_{ij}$ hold in a good
approximation. The magnitude of the deviation from unitarity is connected with the
mass scale of heavy Majorana neutrinos by the approximate relation
$\epsilon^{}_{\alpha\beta}\sim \tilde{\epsilon}^{}_{ij}\sim
\mathcal{O}(M^2_{\rm D}/M^2_{\rm R})$\cite{Xing:2005kh}.
In some viable TeV-scale seesaw models (see, e.g., Ref. \cite{Kersten:2007vk})
one may arrange $M^{}_{\rm D} \sim \mathcal{O}(10^2)~{\rm GeV}$
and $M^{}_{\rm R} \sim \mathcal{O}(10^3)~{\rm GeV}$ to achieve a percent level 
of unitarity-violating effect, although some significant structural cancellations in
$M^{}_\nu \simeq -M^{}_{\rm D} M^{-1}_{\rm R} M^T_{\rm D}$ are unavoidable in
this case. The mass scale of heavy sterile neutrinos in this work is much larger
than the atomic energy transfer, and thus only three light neutrinos can be
produced. In other words, the signature of those heavy degrees of freedom at
low energies is indirectly reflected by the slight departure of the $3\times 3$
PMNS matrix $V$ from unitarity
\footnote{Note that our work is apparently different from the one done in
Ref.~\cite{Dinh:2014toa}, where a light sterile neutrino species of
the ${\cal O}(1)$ eV mass scale as indicated by the short-baseline neutrino
oscillation anomaly has been considered. Such a light sterile neutrino can be
directly generated in the atomic system via its mixing with the active
neutrinos, and hence it violates unitarity of the $3\times 3$ PMNS matrix
in a {\it direct} way.}.
Note that the unitarity-violating parameters $\epsilon^{}_{\alpha\beta}$ and $\tilde{\epsilon}^{}_{ij}$ are not fully independent.
They are connected with each other via the relation
\begin{eqnarray} \label{eq:epsab}
\epsilon^{}_{\alpha\beta} = \sum^{3}_{i=1}\sum^{3}_{j=1} U^{}_{\alpha i}
U^{*}_{\beta j}\tilde{\epsilon}^{}_{ij}\;.
\end{eqnarray}
Given the currently available neutrino oscillation data, precision measurements of
electroweak interactions, and stringent constraints on lepton universality and
lepton flavor violation, it is found that the upper bounds of $|\epsilon^{}_{\alpha\beta}|$
and $|\tilde{\epsilon}^{}_{ij}|$ are at most of order $5\times 10^{-3}$ at the $90\%$
confidence level \cite{Antusch:2006vwa,Antusch:2014woa,Fernandez-Martinez:2016lgt,
Blennow:2016jkn}. If only the neutrino oscillation data are taken into account,
then much looser upper bounds $|\epsilon^{}_{\alpha\beta}| \lesssim \mathcal{O}(0.1)$
and $|\tilde{\epsilon}^{}_{ij}| \lesssim \mathcal{O}(0.1)$ can be achieved \cite{Parke:2015goa}.
The subsequent part of this paper will be devoted to illustrating the effects of
indirect unitarity violation of $V$, as described by $\epsilon^{}_{\alpha\beta}$
and $\tilde{\epsilon}^{}_{ij}$, on the RENP process in an atomic system.

As for the RENP process ${\left| \rm e \right> \rightarrow \left|\rm  g \right> +
\gamma +  \nu^{}_{i}  +  \overline{\nu}^{}_{j} }$, there totally exist six
thresholds in the fine structure of the outgoing photon energy spectrum due to the
finite neutrino masses which are located in the case of vanishing boost
($\bm{p}^{}_{\rm eg}=0$) at the frequencies
\footnote{Note that the threshold frequencies will be altered in
the boosted RENP scenario \cite{Tanaka:2017juo}, which is very interesting for a
further study. In the present work we focus our
attention on putting forward and illustrating our particle-physics idea
by assuming a vanishing boost $\bm{p}^{}_{\rm eg}=0$.}
\begin{eqnarray}
\omega^{}_{ij}  = \frac{E^{}_{\rm eg}}{2} - \frac{\left(m^{}_{i}+m^{}_{j}\right)^2}
{2 E^{}_{\rm eg}} \;.
\end{eqnarray}
One may calculate the rate of such a RENP process with the help of
Eq.~(\ref{eq:Leff2}). An external laser with the frequency $\omega$ can be used to
trigger the transition, and the result for its rate can be factorized
into the expression \cite{Dinh:2012qb,Song:2015xaa,Zhang:2016lqp}
\begin{eqnarray} \label{eq:RENPrate}
\frac{\mathrm{d} N^{}_{\gamma}(\omega) }{\mathrm{d} t} =
6 G^2_{\rm F} V^{}_{\rm tar} n^3 \left(2 J^{}_{p} + 1\right)C^{}_{\rm ep}
\gamma^{}_{\rm vg} \frac{E^{}_{\rm eg}}{E^3_{\rm vg}} I(\omega) \eta^{}_{\omega}(t) \;,
\end{eqnarray}
where $V^{}_{\rm tar}$ represents the target volume, $n$ stands for the number density of
target atoms, $(2 J^{}_{p} + 1)C^{}_{\rm ep}$ denotes the spin factor of the
transition, $E^{}_{\rm vg}$ (or $\gamma^{}_{\rm vg}$) is the energy difference (or the
dipole strength) between the atomic levels
$\left|\rm  v \right>$ and $\left|\rm  g \right>$, and
$\eta^{}_{\omega} (t)$ is the dynamical factor which quantifies
the level of coherence of the medium. The detailed values of these atomic
parameters can be found in Table 9 of Ref.~\cite{Song:2015xaa}.
In the following we shall take the ytterbium (Yb) atomic levels,
for which $E^{}_{\rm eg} = 2.14349~{\rm eV}$ and $E^{}_{\rm vg} = 2.23072~{\rm eV}$,
as an example to show the unitarity-violating effect.
Information about the neutrino properties is hidden in the spectrum function
\begin{eqnarray} \label{eq:I}
I(\omega) = \frac{1}{(\omega-E_{\rm vg})^2}\sum^3_{i=1}\sum^3_{j=1} \Delta_{ij}(\omega)
\left[|a_{ij}|^2 I_{ij}(\omega) - m_im_j {\rm Re}(a^{2}_{ij})\right]
\Theta\left(\omega^{}_{ij}-{\omega}\right) \; ,
\end{eqnarray}
in which the PMNS coefficients $a^{}_{ij} \equiv V^*_{ei} V^{}_{ej} -
(V^{\dagger}V)^{}_{ij}/2$ (for $i, j = 1, 2, 3$) have been
defined below Eq.~(\ref{eq:Leff2}), $\Theta(\omega^{}_{ij} - \omega)$ denotes the
Heaviside function which signifies the kinematic threshold under consideration, and
\begin{eqnarray} \label{eq:Dij}
\Delta_{ij}(\omega)&=&
\frac{\displaystyle\sqrt{\left[E_{\rm eg}\left(E_{\rm eg}-2 \omega\right)
- \left(m_i+m_j\right)^2\right] \left[E_{\rm eg}\left(E_{\rm eg}-2 \omega
\right) - \left(m_i-m_j\right)^2\right]}}
{\displaystyle E_{\rm eg}\left(E_{\rm eg}-2\omega\right)} \;, \hspace{0.4cm}
\nonumber \\
I_{ij}(\omega)&=&
\frac{1}{3}\left[E_{\rm eg}\left(E_{\rm eg}-2\omega\right)+\frac{1}{2}\omega^2
-\frac{1}{6}\omega^2\Delta^2_{ij}(\omega) -\frac{1}{2}\left(m_i^2+m_j^2\right)
\right.
\nonumber \\
&& - \left. \frac{1}{2}\frac{\left(E_{\rm eg}-\omega\right)^2}{E^2_{\rm eg}
\left(E_{\rm eg}-2\omega\right)^2}\left(m_i^2-m_j^2\right)^2 \right] \; .
\end{eqnarray}
Note that the terms proportional to $m^{}_i m^{}_j$ in Eq.~(\ref{eq:I}) exist only
for the Majorana neutrinos, but they are strong suppressed by the smallness of
$m^{}_i$ and $m^{}_j$. These tiny terms will be neglected in the subsequent
discussions, because we are mainly concerned about how the spectrum function
gets distorted under the unitarity violation of $V$ in this work.

The total spectrum in Eq.~(\ref{eq:I}) is linearly composed of the sub-spectra with
six different endpoints, denoted as $\omega^{}_{11}$, $\omega^{}_{12}$, $\omega^{}_{22}$, $\omega^{}_{13}$,
 $\omega^{}_{23}$ and $\omega^{}_{33}$. The location of the stimulating
trigger frequency with respect to those thresholds will be found to be very important
for us to obtain a high sensitivity to the unitarity violation of $V$.
The six thresholds can be classified into three major categories:
$\omega^{}_{\rm I} = (\omega^{}_{11},\omega^{}_{12},\omega^{}_{22})$,
$\omega^{}_{\rm II} = (\omega^{}_{13}, \omega^{}_{23})$
and $\omega^{}_{\rm III} =\omega^{}_{33}$ due to the fact of
$\Delta m^2_{21} \simeq 7.39 \times 10^{-5}~{\rm eV^2} \ll |\Delta m^2_{\rm 31}|
\simeq |\Delta m^2_{\rm 32}| \simeq 2.525 \times 10^{-3}~{\rm eV^2}$ extracted
from a global analysis of current neutrino oscillation data
\cite{deSalas:2017kay,Capozzi:2018ubv,Esteban:2018azc}.
Given the normal ordering (NO) of three neutrino masses with
$m^{}_{1} = 0.05~{\rm eV}$, for example, the energy gap between two different
categories of the thresholds is $\omega^{}_{\rm III} - \omega^{}_{\rm II}
\simeq \omega^{}_{\rm II} - \omega^{}_{\rm I} \sim 10^{-3}~{\rm eV}$,
but the one within the same category (e.g., $\omega^{}_{11} - \omega^{}_{22}$)
is of order $\lesssim 10^{-4}~{\rm eV}$.
\begin{figure}[t!]
	\begin{center}
		\subfigure{
			\hspace{-0.5cm}
			\includegraphics[width=0.45\textwidth]{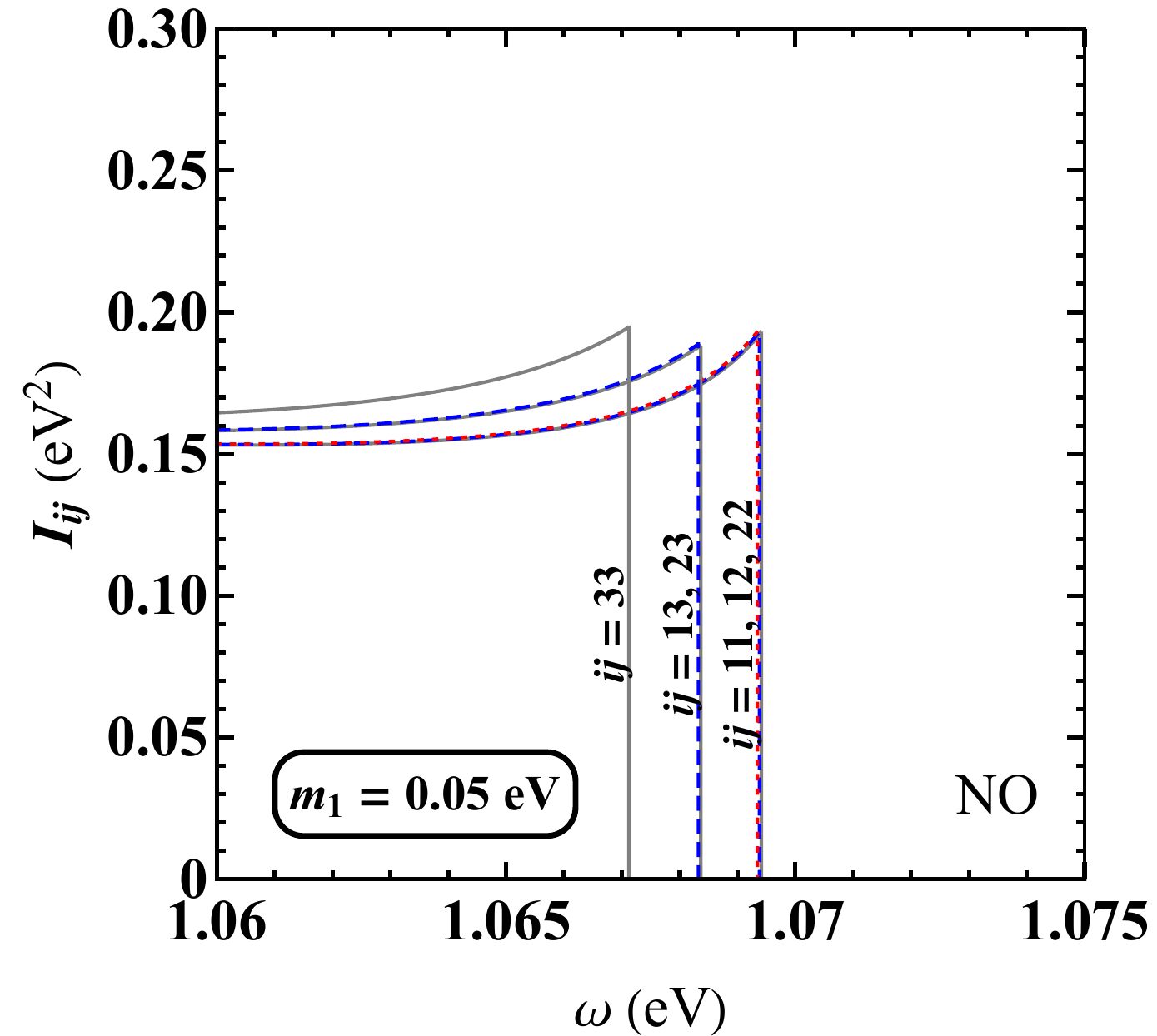} }
		\subfigure{
			\hspace{-0.5cm}
			\includegraphics[width=0.45\textwidth]{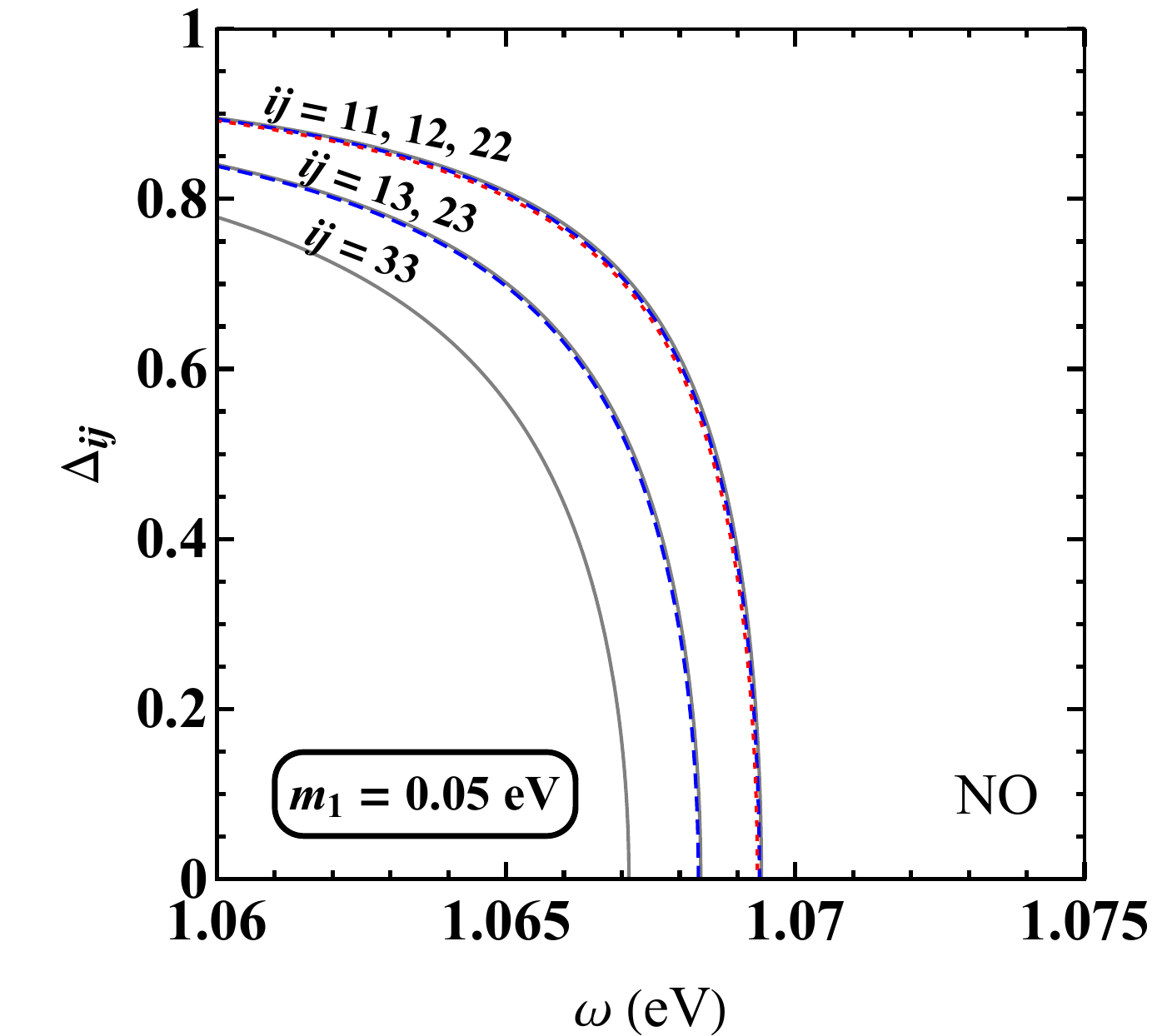} }
	\end{center}
	\vspace{-0.5cm}
	\caption{The functions $I^{}_{ij}(\omega)$ and $\Delta^{}_{ij}(\omega)$
		for different indices ($ij = 11,12,22,13,23,33$), where the normal ordering
		(NO) of three neutrino masses has been taken.
		Different colors have been used to distinguish those curves which almost
		overlap each other.}
	\label{fig:ID}
\end{figure}

The first step of a RENP experiment might be to pin down the absolute neutrino
mass scale --- the value of $m^{}_{1}$ to a reasonable degree of accuracy.
After $m^{}_{1}$ is measured, all the thresholds $\omega^{}_{ij}$ can then be
located in the photon spectrum by inputting the known values of $\Delta m^2_{21}$
and $\Delta m^2_{31}$. On the other hand, the approximate locations of the
thresholds might be directly determined by a rough scanning of the {\it kink}
structure of the photon spectrum \cite{Zhang:2016lqp}. To resolve the gap
between any two different categories of $\omega^{}_{ij}$, the precision of the
laser frequency should be better than the energy gap $\sim 10^{-3}~{\rm eV}$.
Meanwhile, it is more challenging to resolve the tiny gaps within the same
category. We are going to show that one can have a much better sensitivity to
the unitarity violation of $V$ if the laser frequency is chosen to be well
separated from the thresholds inside $\omega^{}_{\rm I}$ or $\omega^{}_{\rm II}$,
such as $\Delta \omega^{}_{\rm I}/(\omega^{}_{{\rm I}, i} - \omega) \ll 1$,
where $\Delta \omega^{}_{\rm I}$ denotes the energy gap inside $\omega^{}_{\rm I}$,
and $\omega^{}_{{\rm I},i}$ is just one of the thresholds within category I.
Under the above condition one may follow a perturbative analysis to verify
the approximate equalities $\Delta^{}_{11} \simeq \Delta^{}_{12} \simeq
\Delta^{}_{22}$ and $I^{}_{11} \simeq I^{}_{12} \simeq I^{}_{22}$ up to a
relative difference of the same order as the small quantity
$\Delta \omega^{}_{\rm I}/(\omega^{}_{{\rm I}, i} - \omega)$.
A similar observation can be achieved for the quantities
associated with $\omega^{}_{\rm II}$. In Fig.~\ref{fig:ID} we plot the functions
$\Delta^{}_{ij}$ and $I^{}_{ij}$ for the NO case with $m^{}_{1} = 0.05~{\rm eV}$,
from which one can see the accuracy of the above approximations.
Since the NO seems to be favored over the inverted ordering (IO) of three
neutrino masses at the $3\sigma$ level
\cite{deSalas:2017kay,Capozzi:2018ubv,Esteban:2018azc}, we only consider the
NO case to numerically illustrate our idea and method in this work
\footnote{In fact, we find that the photon spectrum function has a very similar
behavior in the IO case, and thus we shall not discuss this case in detail
for the sake of simplicity.}.

Given the above conditions for the three thresholds of category $\omega^{}_{\rm I}$,
the emission channels of all the $\nu^{}_{1}$ and $\nu^{}_{2}$ combinations
contribute to the total photon spectrum by an amount of
\begin{eqnarray} \label{eq:111222}
I^{}_{\rm I} & \approx & \frac{1}{\displaystyle\left(\omega-E_{\rm vg}\right)^2}
\Delta_{11}(\omega)I_{11}(\omega) \left(|a_{11}|^2 +2 |a_{12}|^2 + |a_{22}|^2\right)
\nonumber \\
& = & \frac{1}{\displaystyle\left(\omega-E_{\rm vg}\right)^2}
\Delta_{11}(\omega)I_{11}(\omega) \left[\frac{1}{2} -|V^{}_{e3}|^2 + |V^{}_{e3}|^4
+ \left(1-2|V^{}_{e3}|^2\right) \epsilon^{}_{ee} +
\left(\frac{1}{2}-|V^{}_{e1}|^2\right) \tilde{\epsilon}^{}_{11} \right. \hspace{0.2cm}
\nonumber \\
& & + \left. \left(\frac{1}{2} - |V^{}_{e2}|^2 \right) \tilde{\epsilon}^{}_{22} -2
\mathrm{Re}(V^*_{e1} V^{}_{e2} \tilde{\epsilon}^{}_{12}) + \epsilon^2_{ee} +
\frac{\tilde{\epsilon}^2_{11}}{4} + \frac{|\tilde{\epsilon}^{}_{12}|^2}{2} +
\frac{\tilde{\epsilon}^2_{22}}{4} \right] \; .
\end{eqnarray}
If the unitarity-violating parameters $\epsilon^{}_{\alpha\beta}$ and
$\tilde{\epsilon}^{}_{ij}$ are switched off (i.e., $V \to U$), then
$I^{}_{\rm I}$ will be only dependent on the most accurately measured PMNS
matrix element $|U^{}_{e3}|$.
This makes it easier to pin down the unitarity-violating contribution to
$I^{}_{\rm I}$, because this kind of new-physics effect is expected to be
very small and hence easily contaminated by the uncertainties associated
with the PMNS matrix elements.
When the channels with the thresholds $\omega^{}_{13}$ and $\omega^{}_{23}$ are
concerned, the spectrum function receives additional contributions of the form
\begin{eqnarray} \label{eq:1323}
I^{}_{\rm II} & \approx & \frac{2}{\left(\omega-E_{\rm vg}\right)^2}
\Delta_{13}(\omega)I_{13}(\omega) \left(|a_{13}|^2 + |a_{23}|^2\right)
\nonumber \\
& = & \frac{2}{\left(\omega-E_{\rm vg}\right)^2} \Delta_{13}(\omega)I_{13}(\omega)
\left[ |V^{}_{e3}|^2 - |V^{}_{e3}|^4
+ |V^{}_{e3}|^2 \epsilon^{}_{ee} -   \mathrm{Re}(V^{*}_{e1}V^{}_{e3}
\tilde{\epsilon}^{}_{13}) - \mathrm{Re}(V^*_{e2} V^{}_{e3} \tilde{\epsilon}^{}_{23}) \right.
\nonumber \\
&& + \left. \frac{|\tilde{\epsilon}^{}_{13}|^2}{4} +
\frac{|\tilde{\epsilon}^{}_{23}|^2}{4} \right] \;.
\end{eqnarray}
Once again $I^{}_{\rm II}$ will depend only on the matrix element
$|V^{}_{e3}| = |U^{}_{e3}|$ in the unitarity limit.
In particular, all the terms of $I^{}_{\rm II}$, except for
the $\mathcal{O}(\tilde{\epsilon}^2_{ij})$ terms, are suppressed by the smallness of
$|V^{}_{e3}|$. That is why the emission rates of the neutrino pairs
$\nu^{}_{1} + \nu^{}_{3}$ and $\nu^{}_{2} + \nu^{}_{3}$ are insignificant as
compared with the other channels. When the photon energy becomes smaller than
$\omega^{}_{33}$, a contribution of the $\nu^{}_{3}+ \overline{\nu}^{}_{3}$
emission to the photon spectrum reads
\begin{eqnarray} \label{eq:33}
I^{}_{\rm III} & \approx &  \frac{1}{\left(\omega-E_{\rm vg}\right)^2}
\Delta_{33}(\omega)I_{33}(\omega) \left[ \frac{1}{4} -|V^{}_{e3}|^2 +| V^{}_{e3}|^4
+ \left(\frac{1}{2} - |V^{}_{e3}|^2\right) \tilde{\epsilon}^{}_{33} +
\frac{\tilde{\epsilon}^{2}_{33}}{4} \right]\;.
\end{eqnarray}
Furthermore, if the photon frequency is chosen to be far away from all the six thresholds
(e.g., $\omega \rightarrow 0~{\rm eV}$) such that
$\Delta \omega^{}_{}/(\omega^{}_{ij} - \omega) \ll 1$ with $\Delta \omega^{}_{} \lesssim 10^{-3}~{\rm eV}$ denoting the energy difference within the six thresholds,
one will be left with the approximate equalities
$\Delta^{}_{ij} \simeq \Delta^{}_{11}$ and $I^{}_{ij} \simeq I^{}_{11} $
(for $i,j=1,2,3$). In this case the contributions of all the six thresholds can be
summed up as follows:
\begin{eqnarray} \label{eq:Itot}
I^{}_{\rm tot} & \approx &  \frac{1}{\left(\omega-E_{\rm vg}\right)^2}
\Delta_{11}(\omega)I_{11}(\omega) \left[ \frac{3}{4}  + \epsilon^{}_{ee}
+ \frac{1}{2} \sum^{3}_{i=1} \tilde{\epsilon}^{}_{ii} -\sum^{3}_{i=1} \sum^{3}_{j=1}\mathrm{Re}(V^*_{ei}V^{}_{ej}\tilde{\epsilon}^{}_{ij}) \right.
\\ \nonumber
&& + \left. \epsilon^{2}_{ee} + \frac{1}{4} \sum^{3}_{i=1}\sum^{3}_{j=1} |\tilde{\epsilon}^{}_{ij}|^2 \right] \; ,
\end{eqnarray}
in which the leading term is simply a constant, corrected by small
unitarity-violating terms.

The above analytical results tell us that the RENP process is sensitive to the
unitarity-violating parameters $\epsilon^{}_{ee}$, $\tilde{\epsilon}^{}_{11}$, $\tilde{\epsilon}^{}_{22}$, $\tilde{\epsilon}^{}_{12}$,
$\tilde{\epsilon}^{}_{13}$, $\tilde{\epsilon}^{}_{23}$ and $\tilde{\epsilon}^{}_{33}$.
Taking account of Eq.~(\ref{eq:epsab}), we find that $\epsilon^{}_{ee}$ can
actually be expressed as a linear combination
of $\tilde{\epsilon}^{}_{ij}$ (for $i,j=1,2,3$):
\begin{eqnarray} \label{eq:epsee}
\epsilon^{}_{ee} = \sum^{3}_{i=1}\sum^{3}_{j=1}\left|U^{}_{e i}
U^{*}_{e j}\tilde{\epsilon}^{}_{ij} \right| \cos\left(\phi^{}_{i} - \phi^{}_{j} + \phi^{}_{ij}\right) \;,
\end{eqnarray}
where $\phi^{}_{i} \equiv \arg(U^{}_{e i})$ (for $i=1,2,3$) and
$\phi^{}_{ij} \equiv \arg(\tilde{\epsilon}^{}_{ij})$ (for $i,j=1,2,3$).
In the subsequent numerical analysis we shall take $\tilde{\epsilon}^{}_{ij}$
as the original unitarity-violating parameters, and determine the value of
$\epsilon^{}_{ee}$ by specifying the relevant matrix elements of $U$
and $\tilde{\epsilon}$ including their phases. Note that
$V = U \sqrt{ 1 + \tilde{\epsilon}} \simeq U \left(1+\tilde{\epsilon} /2
- \tilde{\epsilon}^2/8\right)$ holds. So the parameters of $U$ should also
be input when calculating the photon spectrum of a RENP process.
\begin{figure}[t!]
	\begin{center}
		\subfigure{
			\hspace{-0.5cm}
			\includegraphics[width=0.45\textwidth]{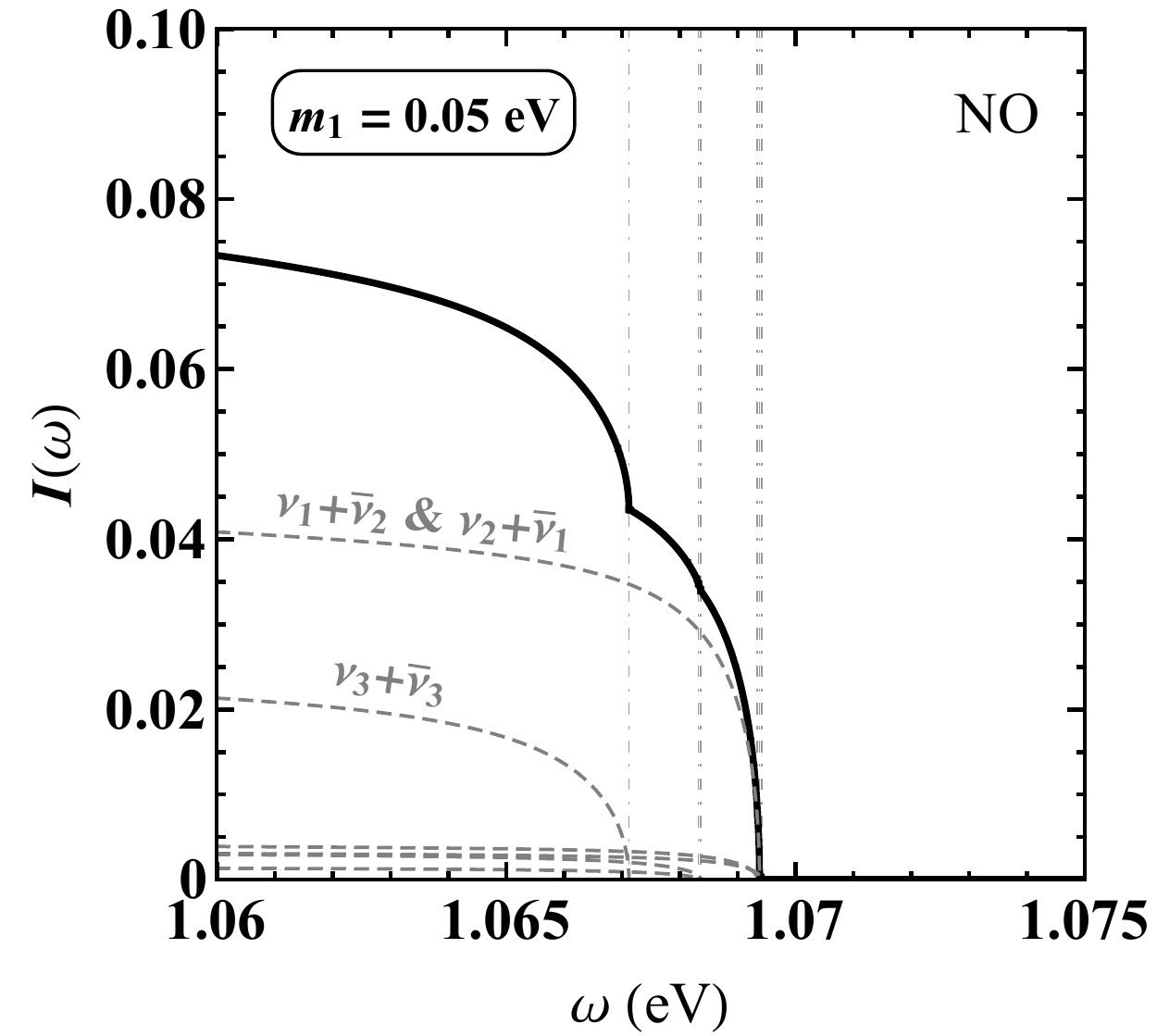} }
		\subfigure{
			\hspace{-0.5cm}
			\includegraphics[width=0.45\textwidth]{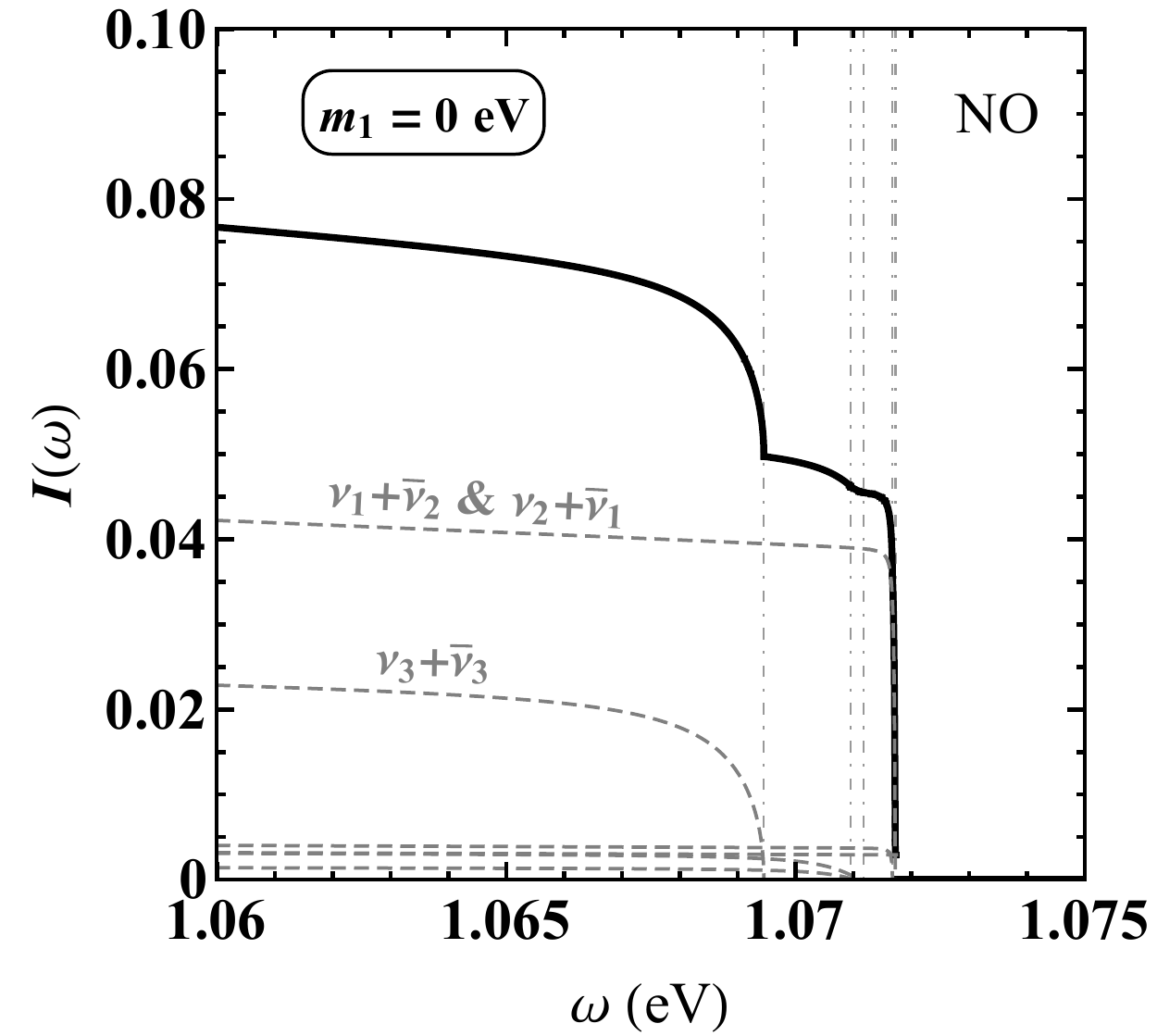} }
	\end{center}
	\vspace{-0.5cm}
	\caption{The photon spectrum function $I(\omega)$ (black solid curves) and
		the contributions of different neutrino pairs (gray dashed curves)
		for $m^{}_{1} = 0.05~{\rm eV}$ (left panel) or $m^{}_{1} = 0~{\rm eV}$
		(right panel), where the normal ordering (NO) of three neutrino masses with
		$\theta^{}_{12} = 33.82^{\circ}$, $\theta^{}_{13} = 8.61^{\circ}$
		and $\theta^{}_{23} =49.6^{\circ}$ of $U$ \cite{Esteban:2018azc} has been taken.
		The vertical dash-dotted lines simply signify all the thresholds in the
spectrum function.}
	\label{fig:specdec}
\end{figure}

We decompose the contributions of different neutrino-pair emissions to the
total photon spectrum with unitarity violation in Fig.~\ref{fig:specdec},
where the solid black curve signifies the total photon spectrum, and the other
curves represent the contributions from explicit neutrino-pair combinations.
The best-fit values of two neutrino mass-squared differences and three flavor mixing
angles of $U$ in the NO case have been taken as the inputs \cite{Esteban:2018azc}:
$\theta^{}_{12} = 33.82^{\circ}$, $\theta^{}_{13} = 8.61^{\circ}$,
$\theta^{}_{23} =49.6^{\circ} $, $\Delta m^{2}_{21} = 7.39 \times 10^{-5}~{\rm eV^2}$
and $\Delta m^{2}_{31} = 2.525 \times 10^{-3}~{\rm eV^2}$.
The vertical dash-dotted lines in Fig. 3 correspond to all the thresholds in the
spectrum function. It is clear that the
emissions of $\nu^{}_{1} + \overline{\nu}^{}_{2}$, $\nu^{}_{2} + \overline{\nu}^{}_{1}$
and $\nu^{}_{3}+\overline{\nu}^{}_{3}$ dominate the total spectrum, and this
observation has already been noticed in Ref.~\cite{Zhang:2016lqp}. Such a result
can be obtained for two simple reasons: (i) the contributions from
$\nu^{}_{1}+\overline{\nu}^{}_{3}$ (or $\nu^{}_{3} + \overline{\nu}^{}_{1}$)
and $\nu^{}_{2}+\overline{\nu}^{}_{3}$ (or $\nu^{}_{3}+\overline{\nu}^{}_{2}$)
are suppressed by the smallness of $|V^{}_{e3}|$, as shown in Eq.~(\ref{eq:1323});
(ii) the emissions of $\nu^{}_{1}+\overline{\nu}^{}_{1}$
and $\nu^{}_{2}+\overline{\nu}^{}_{2}$ are suppressed by the small factors
$(|U^{}_{e1}|^2-1/2)^2 \simeq 0.03$ and
$(|U^{}_{e2}|^2-1/2)^2 \simeq 0.04$.

\begin{figure}[t!]
	\begin{center}
		\subfigure{
			\hspace{-0.5cm}
			\includegraphics[width=0.45\textwidth]{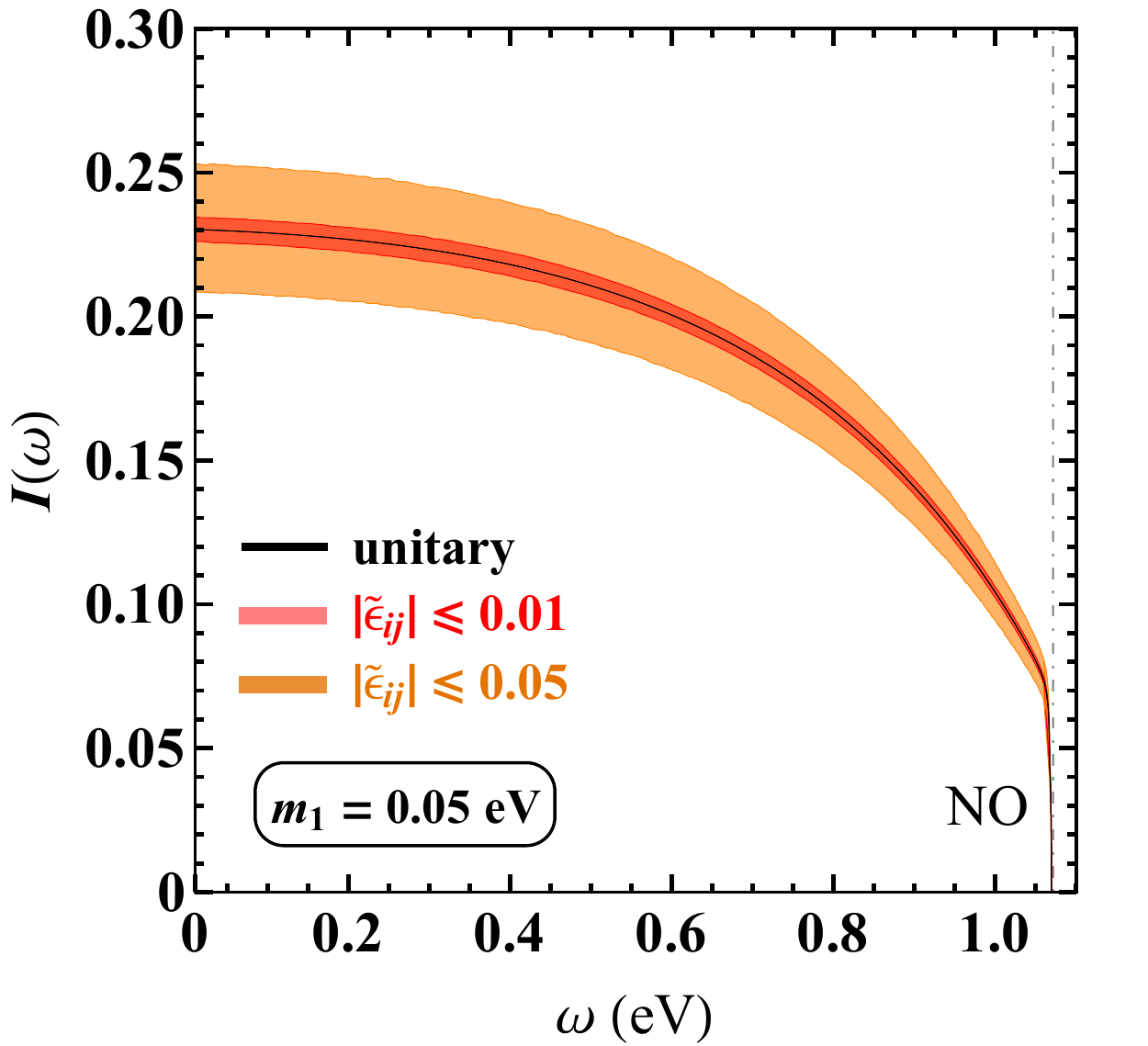} }
		\subfigure{
			\hspace{-0.5cm}
			\includegraphics[width=0.45\textwidth]{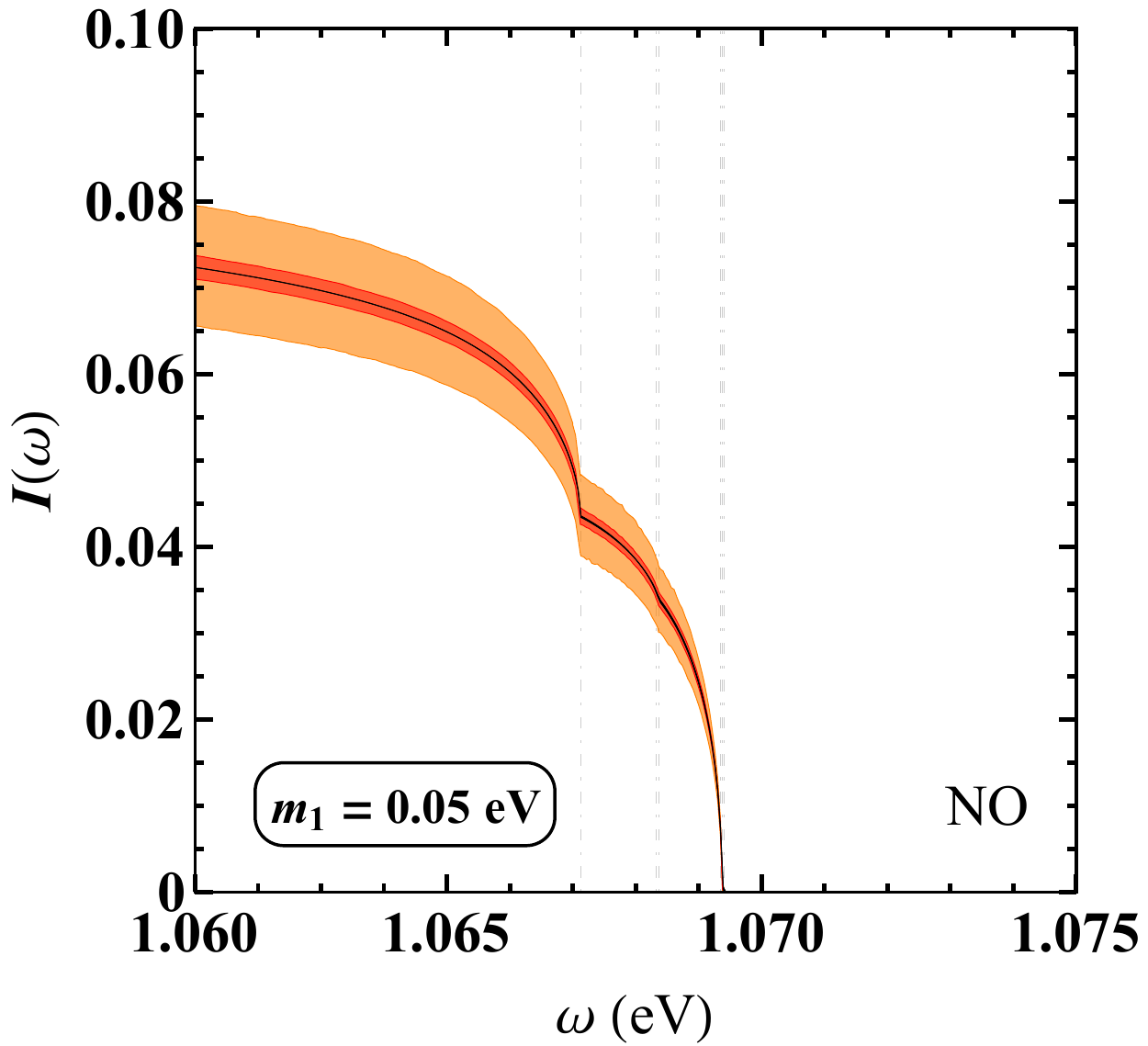} }
		\subfigure{
			\hspace{-0.5cm}
			\includegraphics[width=0.45\textwidth]{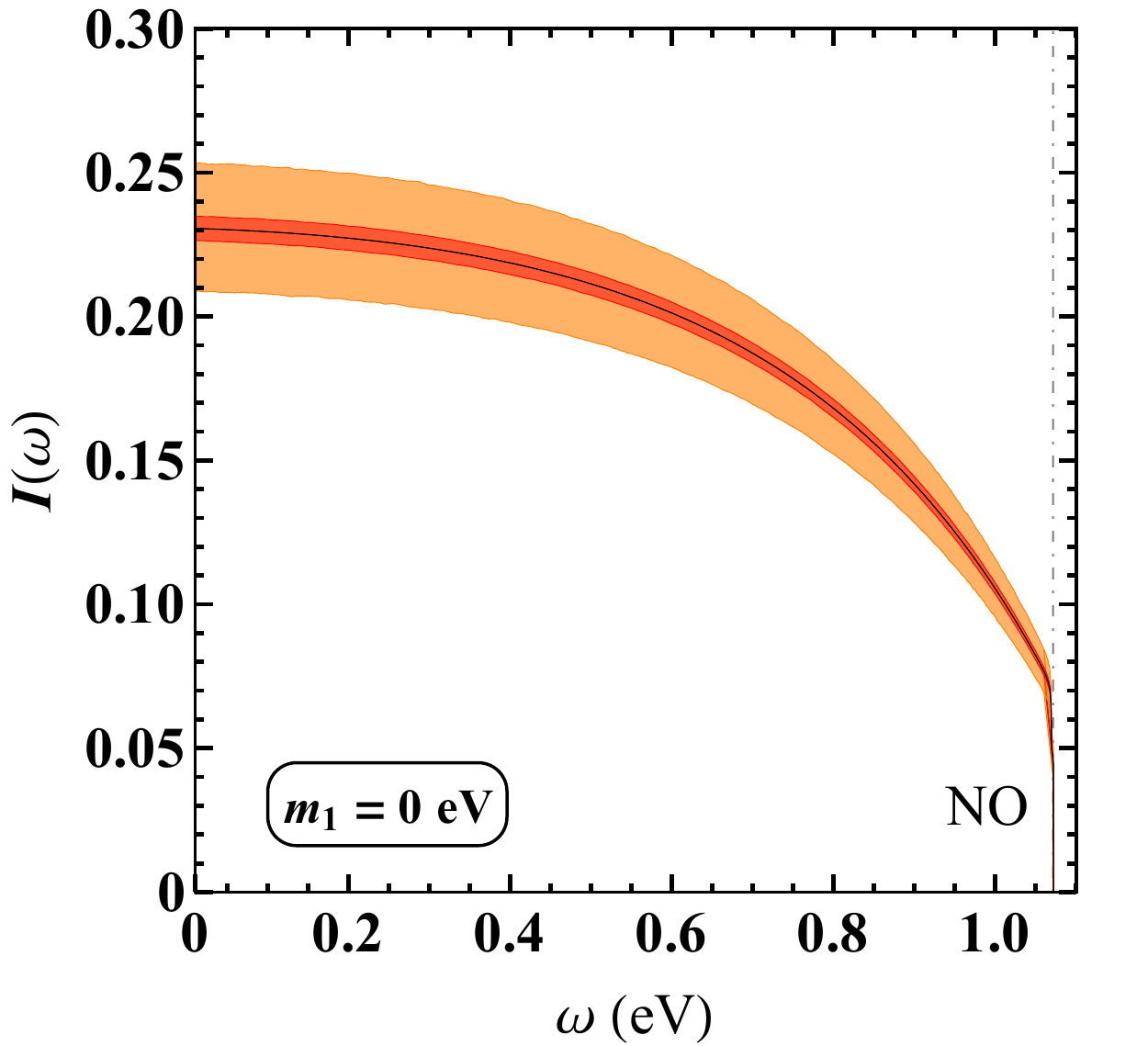} }
		\subfigure{
			\hspace{-0.5cm}
			\includegraphics[width=0.45\textwidth]{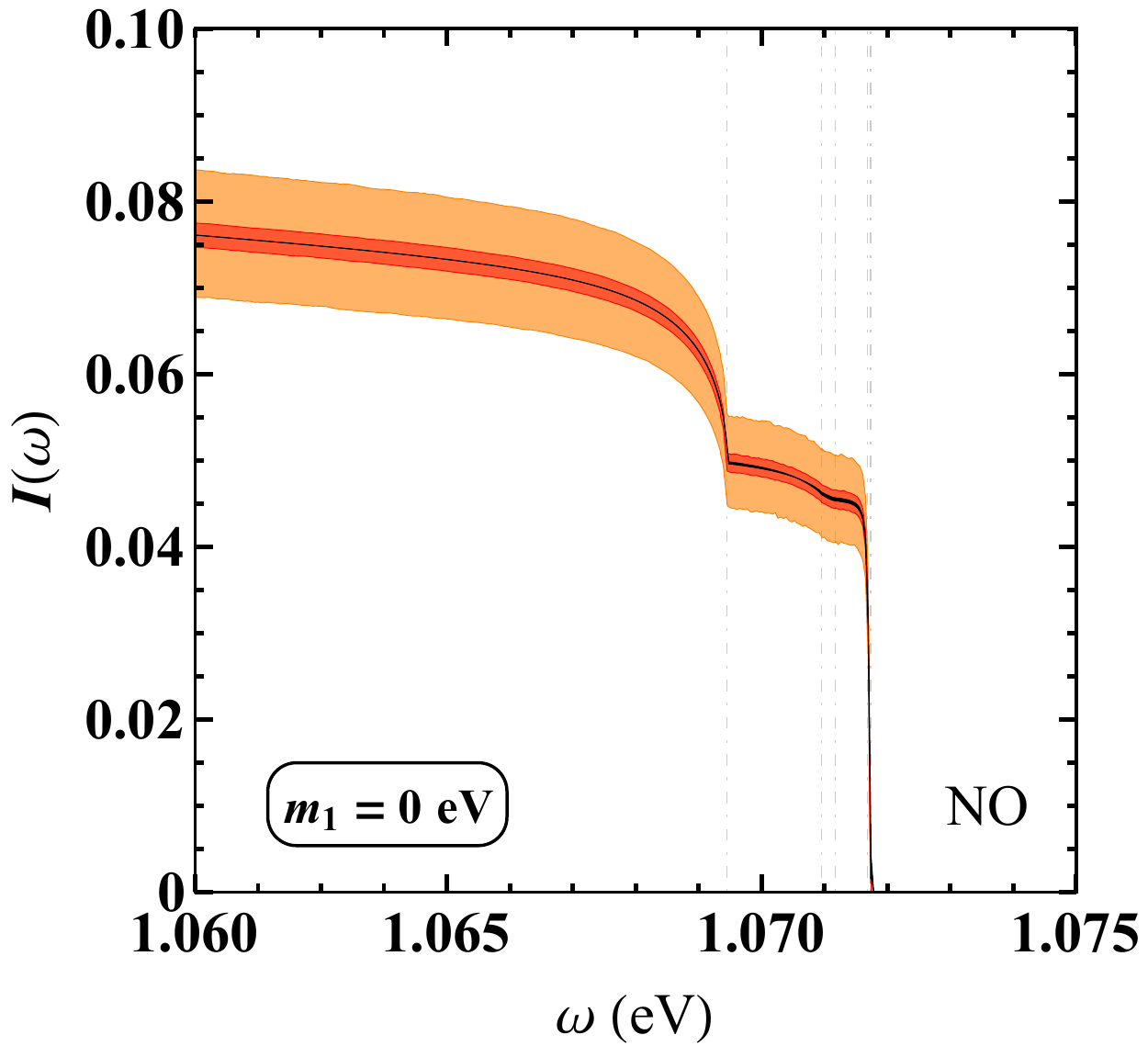} }
	\end{center}
	\vspace{-0.5cm}
	\caption{An illustration of sensitivities of the photon spectrum function
		$I(\omega)$ to the unitarity-violating effects of $V$ in the photon energy
		ranges of $\omega \in [0\cdots 1.2]~{\rm eV}$ (upper-left and lower-left panels)
		and $\omega \in [1.06\cdots 1.075]~{\rm eV}$ (upper-right and lower-right panels).
		The very thin black bands represent the spectra assuming $V$ to be unitary,
		while the much wider orange (or red) bands are produced by allowing $\tilde{\epsilon}^{}_{ij}$ (for $i,j=1,2,3$) to vary
		in the range of $[0 \cdots 0.05]$ (or $[0 \cdots 0.01]$) with arbitrary
phases and by inputting the $3\sigma$ ranges of three flavor mixing angles of $U$
taken from Ref.~\cite{Esteban:2018azc}.
The vertical dash-dotted lines signify all the thresholds in the spectrum function.}
	\label{fig:band}
\end{figure}

Now let us illustrate the overall unitarity-violating effects without assuming any
special values of the relevant parameters. In Fig.~\ref{fig:band}
we require that $|\tilde{\epsilon}^{}_{ij}|$ (for $i,j=1,2,3$) vary
in the range of $[0 \cdots 0.05]$ (orange bands) or $[0\cdots 0.01]$ (red bands),
all the relevant phases of $\tilde{\epsilon}$ and $U$ vary in the range of $[0\cdots 2\pi)$,
and the mixing angles of $U$ vary in their $3\sigma$ ranges as indicated
by the global-fit results \cite{Esteban:2018azc}
(i.e., $\theta^{}_{12} \in [31.61^{\circ}\cdots 36.27^{\circ}]$, $\theta^{}_{13} \in [8.22^{\circ}\cdots 8.99^{\circ}]$ and $\theta^{}_{23} \in [40.3^{\circ}\cdots 52.4^{\circ}]$).
In the left panels of Fig.~\ref{fig:band} the photon spectra
with respect to the whole range of $\omega$ (from $0~{\rm eV}$ to its largest
threshold) have been shown. The upper-left panel stands for the case with
$m^{}_{1} = 0.05~{\rm eV}$, and the lower-left panel corresponds to the case
with $m^{}_{1} = 0~{\rm eV}$. One can see that these two cases are almost
indistinguishable for very small values of $\omega$. This observation is consistent
with Eqs.~(\ref{eq:I}) and (\ref{eq:Dij}). As $\omega \to 0$,
the effect of neutrino masses becomes negligible in comparison with the atomic energy
scale $E^{}_{\rm eg} \simeq 2~{\rm eV} \gg m^{}_i$ for $\rm Yb$.
In the right panel of Fig.~\ref{fig:band} we zoom into the energy region near
the kinematical thresholds, around which one may see some more details.
The standard case without unitarity violation is shown as the much thinner black band.
Appreciable unitarity-violating effects can be observed even
if the uncertainties of all the PMNS neutrino mixing matrix elements
are taken into account.

To quantify the experimental requirement for reaching a given
sensitivity of the unitarity violation of $V$ in measuring the RENP process
for an atomic system, let us follow Refs.~\cite{Song:2015xaa}
and \cite{Zhang:2016lqp} to define the rate normalization factor
\begin{eqnarray} \label{eq:Nnorm}
N^{}_{\rm norm} = \left(\frac{T}{\rm s} \right)
\left(\frac{V^{}_{\rm tar}}{10^2~{\rm cm^3}}\right)
\left(\frac{n}{\rm 10^{21}~{cm^{-3}}} \right)^3 \eta^{}_{\omega} \;.
\end{eqnarray}
The event number can then be determined by using Eq.~(\ref{eq:RENPrate})
as $N^{}_{\rm event} \approx 0.002 \times N^{}_{\rm norm} \times I(\omega)$
for any given observation time $T$ at the frequency $\omega$ and values
of the target volume $V^{}_{\rm tar}$, target number density $n$ and
dynamical factor $\eta^{}_{\omega}$.
To break the degeneracy of effects of the unitarity violation,
the uncertainties of the PMNS matrix elements and the uncertainty of the
experimental parameter $N^{}_{\rm norm}$,
the trigger laser may be set to scan the following frequencies:
$\omega = \left\{ 0.1~{\rm eV}, (\omega^{}_{\rm I} + \omega^{}_{\rm II})/2,
(\omega^{}_{\rm II} + \omega^{}_{\rm III})/2 \right\}$.
The medians of different categories of the thresholds have been chosen to
minimize the uncertainties from the PMNS matrix elements.
An experimental sensitivity to the unitarity-violating parameters
$\tilde{\epsilon}^{}_{ij}$ can be obtained by
minimizing the chi-square function
\begin{eqnarray} \label{eq:chi2}
\chi^2(\tilde{\epsilon}^{}_{ij},\theta^{}_{ij},N^{}_{\rm norm}) =
\chi^2_{\rm osc}(\theta^{}_{ij})+ \sum^{}_{\rm \omega}
\frac{\left[ N^{}_{\rm event}(\tilde{\epsilon}^{}_{ij}) - N^{}_{\rm event}(0)
\right]^2}{N^{}_{\rm event}(0)}
\end{eqnarray}
with respect to $\theta^{}_{ij}$ and $N^{}_{\rm norm}$.
Here $N^{}_{\rm event}(\tilde{\epsilon}^{}_{ij})$ (or $N^{}_{\rm event}(0)$)
stands for the event number with (or without) unitarity violation,
and $\chi^2_{\rm osc}(\theta^{}_{ij})$ includes the experimental information
about the neutrino mixing angles taken from the global-fit results
\cite{Esteban:2018azc}. To roughly reach a $3\sigma$ sensitivity to
$|\tilde{\epsilon}^{}_{ij}| \lesssim \mathcal{O}(0.01)$ (for $i,j=1,2,3$),
which is equivalent to $\Delta \chi^2=9$, we find that
$N^{}_{\rm norm} \gtrsim \mathcal{O}(10^{9})$ is required. Similarly,
$N^{}_{\rm norm} \gtrsim \mathcal{O}(10^{11})$ is needed in order to reach
the $3\sigma$ sensitivity to $|\tilde{\epsilon}^{}_{ij}| \lesssim
\mathcal{O}(10^{-3})$.

\section{Summary}

We have studied the possibility of testing unitarity of the $3\times 3$ PMNS
lepton flavor mixing matrix $V$ in an atomic system with the intriguing RENP process.
The spectrum of the emitted photons will be distorted by the unitarity-violating
effects of $V$. We find that in certain regions of the trigger frequency
only the smallest and best-measured PMNS matrix element $|V^{}_{e3}|$ contributes
to the leading-order term of the transition rate, and in some regions the
leading-order term is even independent of the PMNS matrix elements. This observation
is greatly helpful to enhance the sensitivity of the RENP process to indirect
unitarity violation of $V$. The distortion of the photon spectrum for
the Yb atomic levels has been illustrated by taking into account a reasonable
parameter space.

As the SPAN group is gradually making progress in an experimental realization
of the RENP process \cite{Miyamoto:2014aaa,Miyamoto:2015tva,Miyamoto:2017tva,Hiraki:2018jwu},
the potential to probe or constrain possible unitarity violation of the PMNS
matrix in the atomic system may be very promising in the foreseeable future. The idea
and methodology described here can also be applied to the Raman-stimulated
neutrino pair emission \cite{Hara:2019bur} in a similar atomic system.

We stress that the interplay between atomic physics and particle physics provides
us with a new opportunity to explore new physics hidden at a high energy scale
by using some new techniques at low energies, although this kind of endeavor
is always challenging. Our present work has added a new example in this connection,
to illustrate how to implement an indirect test of the canonical seesaw mechanism by
probing possible unitarity violation of the PMNS matrix in an atomic system. Some
further and more systematic studies along this line of thought will be carried out
later on.

{\it We would like to thank Shun Zhou for some useful discussions.
	This work was supported in part by the National Natural Science
	Foundation of China under grant No. 11775231 and grant No. 11835013
(G.Y.H and Z.Z.X.), and by the JSPS KAKENHI Grant numbers JP 15H02093,
15K13486 (N.S.) and 17H02895 (M.Y.).}


\end{document}